\documentclass{jpsj-suppl}
\usepackage{txfonts} %Please comment out this line unless the txfonts package is availabe in your LaTeX system.
\usepackage{graphicx}
\usepackage{xspace}
\usepackage{sidecap}
\newcommand{\Al}{$^{26}$Al\xspace}
\newcommand{\about}{$\simeq$}

\newcommand{\Fe}{$^{60}$Fe\xspace}
\newcommand{\Co}{$^{56}$Co\xspace} %x

\newcommand{\Ni}{$^{56}$Ni\xspace}
\newcommand{\Ti}{$^{44}$Ti\xspace}

\newcommand{\Msol}{M\ensuremath{_\odot}\xspace}

\title{News from Cosmic Gamma-ray Line Observations}

\author{{Roland \textsc{Diehl}$^{1,2}$}}

\inst{$^{1}$Max Planck Institut f\"ur extraterrestrische Physik, D-85748 Garching, Germany \\ 
$^{2}$Excellence Cluster 'Universe', D-85748 Garching, Germany }

\email{rod@mpe.mpg.de}

\recdate{August 18, 2016}

\abst{The measurement of gamma rays at MeV energies from cosmic radioactivities is one of the key tools for nuclear astrophysics, in its study of nuclear reactions and how they shape objects such as massive stars and supernova explosions.  Additionally, the unique gamma-ray signature from the annihilation of positrons falls into this same astronomical window, and positrons are often produced from radioactive beta decays. Nuclear gamma-ray telescopes face instrumental challenges from penetrating gamma rays and cosmic-ray induced backgrounds. But the astrophysical benefits of such efforts are underlined by the discoveries of nuclear gamma~rays from the brightest of the expected sources. In recent years, both thermonuclear and core-collapse supernova radioactivity gamma~rays have been measured in spectral detail, and complement conventional supernova observations with measurements of origins in deep supernova interiors, from the decay of \Ni, $^{56}$Co, and \Ti. The diffuse afterglow  in gamma rays of radioactivity from massive-star nucleosynthesis is analysed on the large (galactic) scale, with findings important for recycling of matter between successive stellar generations: From \Al gamma-ray line spectroscopy, interstellar cavities and superbubbles have been recognised in their importance for ejecta transport and recycling. Diffuse galactic emissions from radioactivity and positron-annihilation $\gamma$~rays should be connected to nucleosynthesis sources: Recently new light has been shed on this connection, among others though different measurements of radioactive \Fe, and through spectroscopy of positron annihilation gamma~rays from a flaring microquasar and from different parts of our Galaxy.}

\kword{nucleosynthesis, radioactivity, massive stars, supernovae, cosmic rays, microquasar, positrons}

\begin{document}
\maketitle

\section{Introduction}
Nuclear line emission of cosmic origins carries information about nuclear reactions in astrophysical sources. These are fusion reactions creating new nuclei, or the excitation of higher nuclear states from energetic collisions; but also the characteristic emission from positrons as they annihilate with electrons falls into the same astronomical window (see Table 1). This astronomical window is called ``low-energy gamma rays'' ($\gamma$~rays), and characteristically ranges from about 0.1 to 10 MeV.  The two questions that this field strives to answer are: (1) Where and how are new interstellar nuclei produced and released, and (2) what are the trajectories and phase changes of ejecta, from their nucleosynthesis sources into next-generation stars.

Since about 50 years this astronomical window is being studied because of its unique astrophysical information, complementing conclusions about nuclear physics processes in cosmic objects which have been inferred less directly from other astronomical measurements. Large instrumental backgrounds and penetrating radiation combine to make $\gamma$-ray observations a challenging task. The first sky survey was made with the imaging Compton telescope \emph{COMPTEL} \cite{Schoenfelder:1993a}  
aboard the NASA Compton Observatory mission (\emph{CGRO} \cite{Gehrels:1993},
1991-2000), line spectroscopy at high resolution is provided through the \emph{INTEGRAL} \cite{Winkler:2003} 
spectrometer \emph{SPI}  \cite{Vedrenne:2003}
 (since 2002 and likely into the 2020$^{ies}$). 
Diffuse radioactivity from \Al and \Fe has been measured, as well as diffuse emission from positron annihilation, \cite{Jean:2003a,Knodlseder:2005,Bouchet:2010,Siegert:2016}, 
each with affirmations of general understandings of high-energy astrophysics processes, but also with surprises and challenges to astrophysical models (see \cite{Diehl:2013} for a review). Recently, observations of cosmic transient events such as supernova explosions \cite{Diehl:2014,Diehl:2015} and microquasar flares \cite{Siegert:2016a} have added to this.

\section{Cosmic gamma-ray line telescopes}
Cosmic $\gamma$ rays are efficiently absorbed in the Earth atmosphere. Therefore, only space telescopes can be used for such measurements, and cosmic ray bombardment with resulting nuclear excitations of the nuclei within the entire spacecraft instrument materials is unavoidable, leading to large instrumental backgrounds.
Additionally, $\gamma$-ray interactions occur mostly through Compton scattering and pair creation, thus producing energetic secondary particles. Detectors have to be large and massive to capture the entire particle-photon cascade for proper energy measurement. Gamma-ray photons are highly penetrating. Therefore conventional optics to collect and focus the incoming radiation are not feasible, and imaging can only be achieved indirectly (Compton telescopes, or sophisticated shadowing with coded-mask telescopes, earth horizon edge detection, etc.). 

The \emph{CGRO} mission with the imaging Compton telescope instrument \emph{'COMPTEL'} obtained  a sky map for \Al emission and the detection of \Ti $\gamma$ rays from Cas A. \emph{INTEGRAL/SPI} added high resolution spectroscopy of these lines, and also a sky survey in $\gamma$~rays from positron annihilation, and, moreover, the first detections of long-sought \Ni $\gamma$ rays from a supernova of type Ia. \emph{NuSTAR} \cite{Harrison:2013} with its hard X-ray imaging sensitivity up to 80~keV enabled us to obtain an image of Cas~A in \Ti X-ray emission -- a key result to understand core collapse explosions. 

The Compton telescope approach seems now most promising for the next step forward. Although instruments with up to 50 times better sensitivities have been demonstrated and proposed for new missions \cite{Barriere:2006,Greiner:2012,Chiu:2015}, establishing a new observational facility appears unfeasible within the next two decades or more, from budget demands and competing other missions. INTEGRAL, although launched for just a five-year mission in 2002, will be capable to operate well into the 2020 decade, with final de-orbit arranged for 2029. 

\section{Insights on specific source types}

Nuclear $\gamma$~rays result mainly from radioactive isotopes released from nucleosynthesis sites into interstellar space. Explosions such as novae and supernovae are most plausible candidates, but also winds from massive stars are expected to release \Al, for example. Table \ref{line-table} summarises the candidate lines for such measurements \cite{Diehl:2011b}. 
Shorter-lived radioactive isotopes would still be embedded in their sources which are not even transparent to penetrating $\gamma$ rays. Emission from isotopes with decay times above $\sim$100 years may be the superposition from many individual nucleosynthesis events, while for the emission seen from isotopes with shorter decay time a single-source origin can be assumed, even if imaging resolution to few degree precision only is possible. 
Other isotopes, in particular for elements heavier than Fe, are hardly observable, because too low in predicted abundance or too long-lived, both resulting in faint emission. 
Only the brightest of the expected source populations have been seen so far, except for \Al and e$^+$ annihilation $\gamma$ rays, where current data archives cover a range of faint to bright sources and thus already allow refined/detailed astrophysical studies.

%---------------------
\begin{table}[h]
\caption{Radioactive isotopes which are usable for nucleosynthesis studies through $\gamma$-ray lines. Isotopes are  sorted by decay time (column 2); for cases of two-stage decay chains, the decay time for the more short-lived stage is given in brackets.  Column 3 shows the decay chains, column 4 the characteristic $\gamma$-ray line energies (the energy value is in \emph{italics} for lines that have been detected from cosmic sources), and the dominant candidate sources are listed in column 5. }
\label{line-table}
\begin{tabular}{lcccc}
{\bf isotope} & {\bf decay time} & {\bf decay chain} & {\bf energy} & {\bf origin }\\
  & [y] &   &  [keV]  & \\
\hline
$^7$Be & 0.21 & $^7$Be $\rightarrow$ $^{7}$Li$^*$  & 478 & novae \\
$^{56}$Ni & (0.02); 0.304 & $^{56}$Ni $\rightarrow$ $^{56}$Co$^*$ $\rightarrow$ $^{56}$Fe$^*$ + e$^+$  & \emph{158, 812; 847, 1238} & SNe \\
$^{57}$Ni & 1.07 & $^{57}$Ni $\rightarrow$ $^{57}$Co$^*$  & \emph{122} & SNe \\
$^{22}$Na & 3.8 & $^{22}$Na $\rightarrow$ $^{22}$Ne$^*$  + e$^+$  & 1275 & novae \\
$^{44}$Ti & (6~10$^{-4}$); 85 & $^{44}$Ti $\rightarrow$ $^{44}$Sc$^*$  $\rightarrow$ $^{44}$Ca$^*$  + e$^+$  & \emph{68, 78; 1157} & SNe \\
$^{26}$Al & 1.04~10$^6$ & $^{26}$Al $\rightarrow$ $^{26}$Mg$^*$ + e$^+$  & \emph{1809} & SNe, WR,  \\
  &   &   &  &   (novae, AGB) \\
$^{60}$Fe & 3.8~10$^6$; (5.3) & $^{60}$Fe $\rightarrow$ $^{60}$Co$^*$  $\rightarrow$ $^{60}$Ni$^*$  & 59; \emph{1173, 1332} & SNe \\
e$^{+}$ & ($\leq$10$^5$...10$^7$) & e$^+$+H $\rightarrow$ Ps $\rightarrow$ $\gamma\gamma(\gamma)$  & \emph{511, cont} & SNe, novae,  \\
  &   & (or e$^+$+e$^-$  $\rightarrow$ $\gamma\gamma$)  & \emph{511} & compact-object binaries, \\
  &   &   &   & pulsars, etc. \\
  \hline
\end{tabular}
\end{table}
%----------------------------

%%%%%%%%%%%%%%%%%%%%%%%%%%%%%%%%%%%%%%%%%%%%%%%%%%%%%%%%%%%%%%%%%%%%%%%%%%%%%%%
   \begin{SCfigure}  
   \centering
   %\resizebox{\hsize}{!}
   \includegraphics[width=0.5\textwidth]{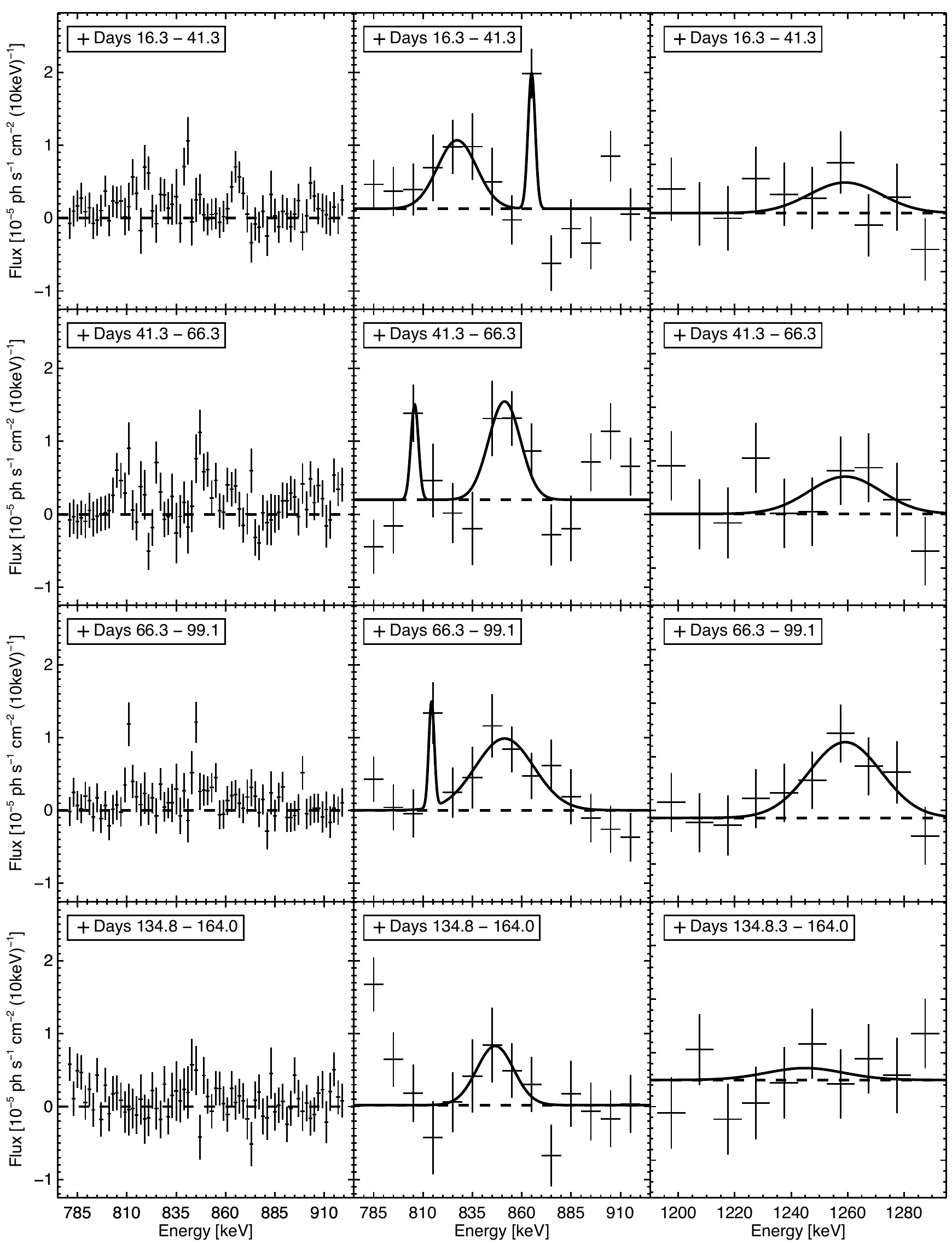}
  \caption{SN2014J $\gamma$ rays from \Co decay. Four successive time intervals are shown, from the time shortly after the explosion \emph{(top)} through maximum brightness 80--100 days later \emph{(third panel/row from top)} to the period of fading brightness due to the 111-day decay time of \Co \emph{(bottom)}. The two rightmost columns show the energy bands of the 847 and 1238 keV lines from \Co decay, respectively, as analysed in broad energy bins; the expected Dopppler-broadened \Co lines can be seen clearly in the brightest epoch (third row from top). The leftmost column shows the band of the 847 keV line (central column) in fine energy bins close to instrumental resolution; some narrow and time variable features appear also part of the \Co emission, in addition to, or altogether composing, the broadened \Co line emission. Note that in all cases, instrumental background is much more intense than the SN2014J countrate; this is reflected in the (1$\sigma$, statistical) error bars, not getting smaller for broad-bin analysis as broader bins include more background counts. }
   \label{Fig_Co_SN2014J}%
   \end{SCfigure}
%%%%%%%%%%%%%%%%%%%%%%%%%%%%%%%%%%%%%%%%%%%%%%%%%%%%%%%%%%%%%%%%%%%%%%%%%%%%%%%%

\subsection{Thermonuclear supernovae}
Supernovae of type Ia have long been considered the best cases to confirm the radioactive decay chain originating from \Ni being the power source of supernova light, through nuclear $\gamma$-ray line spectroscopy \cite{Clayton:1969}. 
Lately, SN2014J At a distance of 3.5~Mpc was sufficiently nearby to obtain such a direct measurement through nuclear $\gamma$-ray line spectroscopy, and significant results could be obtained with INTEGRAL \cite{Diehl:2014,Churazov:2014,Diehl:2015,Churazov:2015,Isern:2016}. Confirming the general supernova model at first glance \cite{Churazov:2014}, the details of the results also present new puzzles: The expected lines from $^{56}$Co decay at 847 and 1238 keV were seen brightest after about three months, and Doppler broadened with typical supernova ejecta velocities. But additionally, early data showed \Ni lines at 158 and 812 keV at a time where a centrally-ignited white dwarf would still occult its inner \Ni products; a surface explosion was then proposed to have triggered the supernova \cite{Diehl:2014,Isern:2016}. Also, the gradual appearance of $^{56}$Co lines and their inferred light curve does not quite match any available model, with observed irregularity pointing at a complex 3D morphology of \Ni ejecta and overlying, absorbing, materials \cite{Diehl:2015}. 
%Unfortunately, the next best candidate event, SN2011fe, could not be detected in $\gamma$~rays \cite{Isern:2013}. 
Gamma-ray spectra thus have demonstrated their complementing, and partially unique, diagnostic contributions to unravel the characteristics of thermonuclear explosions, with this nearby supernova. INTEGRAL's remaining operational lifetime should provide more sufficiently-nearby such opportunities. 

\subsection{Core collapse supernovae}
The NuSTAR image of the 350 year old supernova remnant Cas A at 3.4 kpc distance is a key result for understanding core collapse explosions \cite{Grefenstette:2014}.
%, demonstrating the new and unique information from nuclear emission: 
%The the X-ray recombination line from Fe, which was measured and imaged by Chandra, depends on the ionization state of Fe, in addition to its abundance. The comparison of the NuSTAR \Ti and the Chandra Fe line image of Cas A makes evident the risk of premature or simplified conclusions. % when density effects or ionization state uncertainties become relevant. 
%But atomic and nuclear data also complement each other, if properly interpreted, and the inner explosion characteristics can be well constrained even 350 years after the supernova. 
Clumpiness and 3D effects are characteristic for the Cas A explosion. Results from SN1987A point into the same direction \cite{Kjaer:2010,Boggs:2015}. But both these supernovae may not be representative of core collapse explosions in general: among other arguments, $\gamma$-ray surveys do not find as many \Ti sources  \cite{The:2006} as inferred from the Galactic supernova rate constraint based on \Al from massive stars (see below). Apparently, \Ti ejection occurs for a somewhat rare subclass of core collapse supernovae. This is consistent with findings from numerical supernova modeling, as also reported at this conference. 
Note that a radioactive decay cascade such as $^{44}$Ti-$^{44}$Sc-$^{44}$Ca allows consistency checks of both abundance and kinematics information about $^{44}$Ti ejecta; the recent comparison of 68/78~keV and 1157~keV line parameters with INTEGRAL/SPI \cite{Siegert:2015} indicates some tension. This may open a path towards measuring the acceleration of cosmic rays in young supernova remnants through nuclear-line emissions.

\section{Diffuse radioactivities}
\subsection{\Al throughout the Galaxy}
\begin{SCfigure}
   \centering
   %\resizebox{\hsize}{!}
   \includegraphics[width=0.4\textwidth]{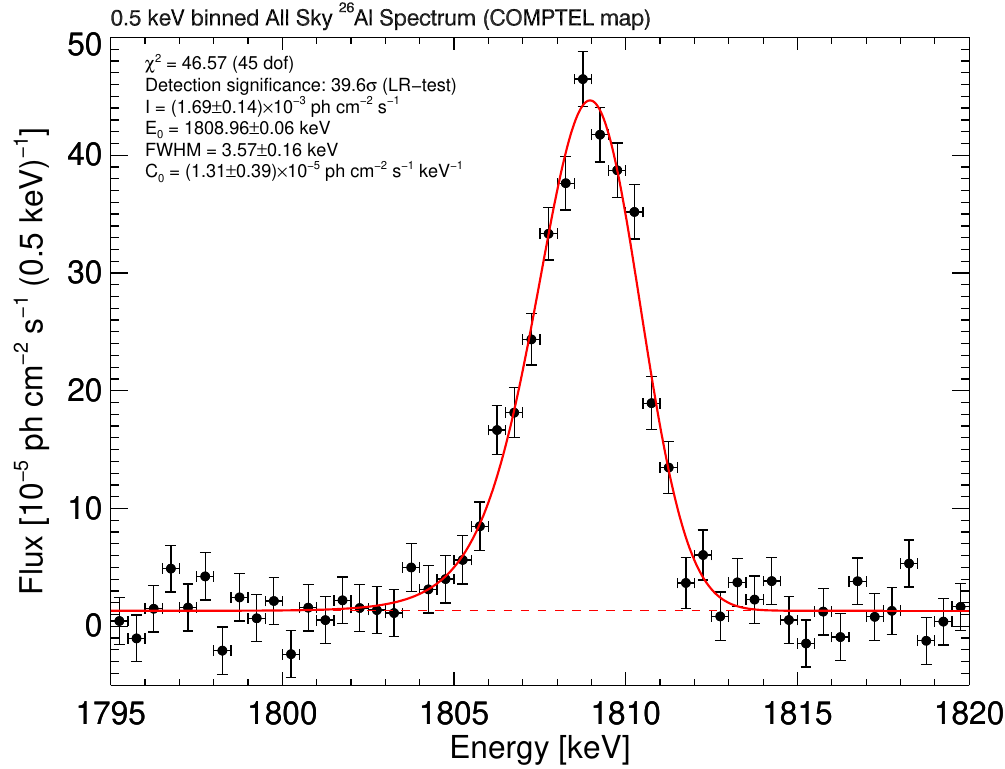}
 \caption{The INTEGRAL/SPI $\gamma$-ray spectrum from \Al decay in interstellar gas  ($\tau\sim$1~Myr). Cumulative all-sky measurements over 14 years obtain a high quality spectrum representing \Al throughout our Galaxy, and resolve the line shape in detail.}
   \label{Fig_26Al}%
   \end{SCfigure}

\Al radioactivity ($\tau\sim$1~My) and the observation of its characteristic $\gamma$-ray line at 1808.73~keV \cite{Mahoney:1984,Diehl:2006c}  has been understood as one of the most direct proofs that nucleosynthesis of \Al is ongoing in our Galaxy after more than ten Gyrs of evolution, and probably in a steady state. The most recent spectrum from 14 years of cumulative SPI single-detector events, analysed with a high spectral resolution background method, is shown in Fig.~\ref{Fig_26Al} (left) \cite{Diehl:2017}. Earlier measurement of the characteristic systematic Doppler shifts of the line with Galactic longitude \cite{Diehl:2006d,Kretschmer:2013} had shown that the observed \Al $\gamma$ rays originate from sources throughout the Galaxy, including its distant and otherwise occulted regions at and beyond the inner spiral arms and bulge; the shifts of the \Al-line centroid energy are due to the Doppler effect from large-scale Galactic rotation. 

Comparing measured Doppler shifts per longitude between \Al $\gamma$~rays and other sources (such as molecular clouds seen in CO or masers related to young stars) have led to a remarkable insight: The apparent \Al velocities exceed expected object motion Doppler shift by about 200~km~s$^{-1}$, which led to the interpretation that \Al is released into low-density cavities blown by stellar winds and supernova in massive star groups located at the leading edges of spiral arms \cite{Kretschmer:2013,Krause:2015}. Then, cavities will be asymmetric in extent around these sources, and prefer forward motion in the large-scale Galactic rotation, so that the systematically higher \Al derived velocities can be understood. More quantitatively, such superbubbles will have an extent into the inter-arm regions of typically \about~100~pc or more, while the massive star groups themselves have moved away from spiral arms within several Myrs only by tens of pc. Massive star ejecta then can be understood to first appear as hot plasma in superbubbles, then stream into the inter-arm medium and the halo of our Galaxy. This interpretation also is supported by views of external spiral galaxies, where such forward-edge bias of massive star and interstellar cavity observations can be seen within the co-rotation radius. It remains to be assessed, how this may delay the incorporation of new nuclei into next generation stars, and thus the chemical enrichment dynamics within a spiral galaxy. 

The all-sky-integrated measured \Al flux gives an estimate for the Galactic core collapse supernova rate \cite{Diehl:2006d}, using first-order Galaxy models such as exponential disks, and combining these with \Al yields for massive stars and their supernovae with an appropriate weighting due to the stellar mass distribution function (IMF).  
This includes different systematics and uncertainties than previous measurements of the Galaxy's core-collapse supernova rate, and yet confirms the other values inferred for this massive star activity. 
A possible bias when attributing the observed flux to such a Galaxy-wide interpretation are nearby groups of massive stars. The inclusion of the identified specific regions of Sco-Cen, Cygnus, and Orion at their known distances revises the total Galactic \Al mass estimate now to 2.0~\Msol. with a $\pm$0.3~\Msol statistical uncertainty. The inferred rate of Galactic core-collapse supernovae is correspondingly somewhat reduced to 1.3$\pm$0.4 events per century.

%%%%%%%%%%%%%%%%%%%%%%%%%%%%%%%%%%%%%%%%%%%%%%%%%%%%%%%%%%%%%%%%%%%%%%%%%%%%%%%
%%%%%%%%%%%%%%%%%%%%%%%%%%%%%%%%%%%%%%%%%%%%%%%%%%%%%%%%%%%%%%%%%%%%%%%%%%%%%%%%

\subsection{Positron annihilation gamma rays}
The large scale distribution of characteristic $\gamma$~rays from positron annihilation has been an astrophysics puzzle \cite{Prantzos:2011}, as a bright and extended emission region in the central region of our Galaxy dominates the 511~keV sky \cite{Jean:2003a,Knodlseder:2005}, while plausible sources of interstellar positrons should follow a distribution extended along the disk of the Galaxy. Recently, the 13-year database of INTEGRAL measurements had been used to shed new light on the large scale models of Galactic positron annihilation distributions \cite{Siegert:2016}. High-quality spectra for the bright bulge, but also for the Galactic disk, and for a newly discriminated central source  have ben determined. Although all spectra show a modestly-broadened 511 keV line together with a continuum from 3-photon annihilation through the intermediate formation of a positronium atom, the positronium contribution may be somewhat reduced in the disk compared to the bulge. The new central source spectrum is interesting in spite of its modest significance of \about~3$\sigma$, because the indications of a larger line broadening and small red shift would fit expectations from an origin in the central molecular zone or the vicinity of Sgr~A. 

Candidate positron sources that have been claimed or proposed  include radioactive decays from the nucleosynthesis source variety (novae, massive stars and core collapse supernovae, i.e. from \Ti, \Al), and thermonuclear supernovae (\Ni), but also positron origins from pulsars and their high-energy particle cascades, as well as from accreting compact objects in binary systems with presumed jet-like pair plasma ejections, and even dark matter.
% had been proposed to be a source of positrons, inspired by the unusual brightness of the bulge region. The spatial distribution expected from dark matter annihilation seems to reproduce the observed emission profile remarkably well \cite{Skinner:2014}. 
But a recent search for 511 keV emission from nearby dwarf spheroidal galaxies (which should be dominated by dark matter) does not match with such expectations \cite{Siegert:2016b}, disfavoring a dark matter origin. 
Also interesting is a recent measurement of kinematically broadened annihilation signatures from a flaring microquasar: The V404 Cygni source showed  an exceptionally bright outburst in June 2015, and for the first time, significant excess emission above the Comptonization spectrum, which is characteristic for such sources, could be shown \cite{Siegert:2016a} as plausibly consistent with the presence of pair plasma near the compact source. This had been presumed in theoretical models to characterise the release of gravitational energy near the black hole horizon \cite{Remillard:2006,Laurent:2012}. 

\section{Summary and Conclusions}
Measurements of high energy photons originating from characteristic nuclear de-excitations or from annihilation of positrons are an important contribution to multi-messenger astronomy. Nuclear fusion reactions inside stars and stellar explosions produce characteristic radioactive isotopes, which then reflect the conditions inside their sources through their abundance and kinematics. 
The measurements of $\gamma$~rays in the nuclear window are rich in information about physical processes that are otherwise hard to get by; it is desirable to have a clear perspective for the future of such observations.   

%\section*{Acknowledgments}
{\small \noindent{\bf Acknowledgments}.
We appreciate the support from ASI, CEA, CNES, DLR, ESA, INTA, NASA and OSTC of the INTEGRAL ESA space science mission with the SPI spectrometer instrument project. 
R.D. is grateful for special support from a NAOJ fellowship. 
This work was also supported from the Munich cluster of excellence \emph{Origin and Evolution of the Universe}.}

%%%%  References %%%%%

%%%%%%%%%%%%%%%%%%%%%%%%%%%%%%%%% satisfy BibTeX from ADS %%%%%%%%%%%%%%

          % Astronomical Journal
\newcommand{\actaa}{Acta Astron. }%
  % Acta Astronomica
\newcommand{\araa}{Ann.Rev.Astron.\&Astroph. }%
          % Annual Review of Astron and Astrophys
\newcommand{\apj}{Astroph.J. }%
          % Astrophysical Journal
\newcommand{\apjl}{Astroph.J.Lett. }%
          % Astrophysical Journal, Letters
\newcommand{\apjs}{Astroph.J.Supp. }%
          % Astrophysical Journal, Supplement
\newcommand{\ao}{Appl.~Opt. }%
          % Applied Optics
\newcommand{\apss}{Astroph.J.\&Sp.Sci. }%
          % Astrophysics and Space Science
\newcommand{\aap}{Astron.\&Astroph. }%
          % Astronomy and Astrophysics
\newcommand{\aapr}{Astron.\&Astroph.~Rev. }%
          % Astronomy and Astrophysics Reviews
\newcommand{\aaps}{Astron.\&Astroph.~Suppl. }%
          % Astronomy and Astrophysics, Supplement
\newcommand{\aj}{Astron.Journ. }%
          % Astronomical Journal
\newcommand{\azh}{AZh }%
          % Astronomicheskii Zhurnal
\newcommand{\memras}{MmRAS }%
          % Memoirs of the RAS
\newcommand{\mnras}{Mon.Not.Royal~Astr.~Soc. }%
          % Monthly Notices of the RAS
\newcommand{\na}{New Astron. }%
  % New Astronomy
\newcommand{\nar}{New Astron. Rev. }%
  % New Astronomy Review
\newcommand{\pra}{Phys.~Rev.~A }%
          % Physical Review A: General Physics
\newcommand{\prb}{Phys.~Rev.~B }%
          % Physical Review B: Solid State
\newcommand{\prc}{Phys.~Rev.~C }%
          % Physical Review C
\newcommand{\prd}{Phys.~Rev.~D }%
          % Physical Review D
\newcommand{\pre}{Phys.~Rev.~E }%
          % Physical Review E
\newcommand{\prl}{Phys.~Rev.~Lett. }%
          % Physical Review Letters
\newcommand{\pasa}{PASA }%
  % Publications of the Astron. Soc. of Australia
\newcommand{\pasp}{Proc.Astr.Soc.Pac. }%
          % Publications of the ASP
\newcommand{\pasj}{Proc.Astr.Soc.Jap. }%
          % Publications of the ASJ
\newcommand{\rpp}{Rep.Prog.Phys. }%
          % Publications of the ASJ
\newcommand{\skytel}{Sky\&Tel. }%
          % Sky and Telescope
\newcommand{\solphys}{Sol.~Phys. }%
          % Solar Physics
\newcommand{\sovast}{Soviet~Ast. }%
          % Soviet Astronomy
\newcommand{\ssr}{Space~Sci.~Rev. }%
          % Space Science Reviews
\newcommand{\nat}{Nature }%
          % Nature
\newcommand{\iaucirc}{IAU~Circ. }%
          % IAU Cirulars
\newcommand{\aplett}{Astrophys.~Lett. }%
          % Astrophysics Letters and Communications
\newcommand{\apspr}{Astrophys.~Space~Phys.~Res. }%
          % Astrophysics Space Physics Research
\newcommand{\nphysa}{Nucl.~Phys.~A }%
          % Nuclear Physics A
\newcommand{\physrep}{Phys.~Rep. }%
          % Physics Reports
\newcommand{\procspie}{Proc.~SPIE }%
          % Proceedings of the SPIE
         
\newcommand\newblock{}

\bibliographystyle{aip} 
%\bibliography{rod-refs16}

\end{document}